# Equatorwards Expansion of Unperturbed, High-Latitude Fast Solar Wind

G.D. Dorrian[1, 2], A.R. Breen[2,†], R.A. Fallows[2,3], and M.M. Bisi[2].

[1]*Institute of Astronomy and Astrophysics, National Observatory of Athens, Greece.*

[2]*Institute of Mathematics and Physics, Aberystwyth University, Wales, UK.*

[3]*ASTRON, The Netherlands Institute for Radio Astronomy, Dwingeloo, The Netherlands.*

[†]*Deceased.*

**Abstract** We use dual-site radio observations of interplanetary scintillation (IPS) with extremely long baselines (ELB) to examine meridional flow characteristics of the ambient fast solar wind at plane-of-sky heliocentric distances of 24-85 solar radii ($R_\odot$). Our results demonstrate an equatorwards deviation of 3-4° in the bulk fast solar wind flow direction over both northern and southern solar hemispheres during different times in the declining phase of Solar Cycle 23.

## 1. Introduction

The rapid variation of the apparent intensity of distant, small-diameter radio sources was first observed in the early 1960s and it was soon realised that observations of this interplanetary scintillation (IPS) could provide information on the solar wind (Hewish, Scott, and Wills, 1964; Dennison and Hewish, 1967). Density irregularities in the solar wind induce phase variations along the raypath of a radio beam as it passes close to the Sun, leading to rapid variations in the amplitude recorded at an Earth based receiver. Developments in the technique led to the use of two-station observations, in which the scintillation pattern is measured at two widely separated receivers (Armstrong and Coles, 1972). Given that the distance to the radio source is much greater than the distance between the receivers, the two raypaths can be assumed to be parallel. In such a case the correlation between the signals detected at the two sites should be a maximum when the projection into the plane of the sky (henceforth referred to as "PoS") of the baseline separation between each raypath is parallel to the PoS solar wind outflow (Bourgois *et al*., 1985; Coles, 1995; Grall *et al*., 1996).

Observations made as the observational geometry changes with the rotation of the Earth can thus be used to investigate non-radial expansion of the solar wind in the meridional direction (*e.g.* Moran *et al*., 1998), with the sensitivity to such meridional flows improving as the baseline lengths for the observations increase. Considerations of the geometry for such observations showed that any deviations from radial flow detected are overwhelmingly likely to arise from meridional flow, as implausibly high azimuthal velocities would be required to produce any detectable effect (Moran *et al*., 1998).

The *European Incoherent SCATter* radar facility (EISCAT: Rishbeth and Williams, 1985) employs a series of parabolic antennas, and is used primarily for ionospheric studies. The large separation between the constituent antennas (up to 390 km, for the mainland sites) does, however, make the EISCAT facility ideal for use with observations of IPS, and it has been used in this role in numerous previous studies (*e.g.* Bourgois *et al*., 1985; Breen *et al*., 1996; Breen *et al*., 2002; Fallows, Williams, and Breen, 2002). Comprehensive descriptions of the use of EISCAT for observations of IPS, including dual-frequency observations, can be found in Fallows *et al*. (2006) and Fallows, Breen, and Dorrian (2008).

More recent studies combining observations from EISCAT and the *Multi-Element Radio Link* (MERLIN: Thomasson, 1986) began in 2002 resulting in the availability of baseline lengths of over 2000 km (*e.g.* Bisi *et al*., 2005, 2007; Breen *et al*., 2006, 2008). These extremely-long baseline (ELB) observations of IPS dramatically improve our ability to resolve small variations in solar wind outflow speed and its meridional direction.

## 2. Observations

The observations used in this paper were made using a combination of both the MERLIN and EISCAT antennas. The MERLIN sites at Jodrell Bank, Knockin, and Cambridge across England, and the EISCAT sites at Kiruna (Sweden) and Sodankylä (Finland) made observations of IPS centred on 1420 MHz, while the EISCAT Tromsø (Norway) antenna observed at a central frequency of 928 MHz; Fallows *et al*. (2006) have shown that reliable results can be obtained using multiple frequencies.

We discuss five observations of the fast solar wind at high heliographic latitudes between 2002 and 2006. The primary aim of the study was to investigate whether any non-radial flow in the fast solar wind was detectable using ELB IPS. Limitations on the time available on MERLIN restricted the number of ELB observations, but three high-quality sets of observations were made of scintillation in the northern polar fast stream between 2002 and 2005, with more observations of scintillation at high southern latitudes in 2006.

The northern hemisphere observations in 2002, 2004, and 2005 took advantage of the strong source J0319+415 (3C84) and are described, together with the results of a preliminary analysis, in Bisi *et al*. (2005) and Bisi (2006). The latitude of the point of closest approach of the raypath to the Sun (the P-point) ranged from 56°-60°N, while the 2006 campaign added observations of J0521+166 (3C138), where the P-Point lay at 81°S. The heliocentric distances of the P-Point throughout all campaigns ranged from 24 $R_\odot$ to 85 $R_\odot$. Each observation lasted several hours so that, due to the Earth's rotation, a significant change in perpendicular baseline occurred for as many receiver-to-receiver baseline combinations as possible. The key parameters for the observations are summarised in Table 1.

| Date | Radio Source | Latitude of P-Point (heliographic) | Heliocentric distance of P-Point ($R_\odot$) | Solar Wind PoS Velocity at maximum cross-correlation (km s$^{-1}$) |
|---|---|---|---|---|
| 15 May 2002 | J0319+415 | 60.0°N | 82.9 | 800 |
| 12 May 2004 | J0319+415 | 56.2°N | 84.7 | 785 |
| 13 May 2005 | J0319+415 | 57.4°N | 84.1 | 733 |
| 10 June 2006 | J0521+166 | 82.5°S | 24.9 | 791 |
| 11 June 2006 | J0521+166 | 80.9°S | 24.3 | 842 |

**Table 1.** Summary of the ELB observations of IPS discussed in this paper. The P-Point is the point of closest approach of the IPS raypath to the Sun. Included is the PoS velocity for each solar wind stream, confirming that it is the fast solar wind which is being sampled in each case.

Figures 1, 2 and 3, show the raypath for each observation run, ballistically back-projected onto the corona down to an altitude of 2.5 R$_\odot$. The technique of observing IPS is preferentially sensitive to scattering events along the raypath that are close to the P-Point. It is assumed (*e.g.* Breen *et al*., 2002; Bisi, 2006) that rarefied coronal structures correspond to fast solar wind flow, and this is confirmed by the observed solar wind velocities shown in Table 1, for each observation. In all cases the observed scattering along the raypath is dominated by high-latitude fast solar wind streams.

The raypath for the observation of J0319+415 on 15 May 2002 lay mainly above the northern polar coronal hole and can therefore be assumed to be mostly immersed in fast solar wind (Figure 1a). The ends of the raypath lay above bright corona, suggesting that they were immersed in slow wind – and the cross-correlation functions for the observation did indeed show clear fast and slow peaks (Bisi *et al*., 2005; Breen *et al*., 2006; Bisi, 2006). The two peaks were, however, well-separated and there was no evidence for the fast stream being anything other than unperturbed fast wind from the northern polar hole.

On 12 May 2004 the raypath for the observation of J0319+415 was again mainly above the northern polar hole (Figure 1b) and this observation can be interpreted as being dominated by unperturbed fast flow from moderately-high northern latitudes. The raypaths for the observations of J0319+415 on 13 May 2005 lay predominantly above the northern polar coronal hole (Figure 2), but a portion of the raypath overlay a latitudinally-extended boundary region between the coronal hole and a dense coronal structure. It is therefore likely that a SIR would be present in this part of the raypath. However it is important to note that the P-Point lies approximately 65° Earthward of the SIR boundary and is immersed in rarefied coronal plasma, indicating the presence of ambient fast solar wind crossing this portion of the raypath.

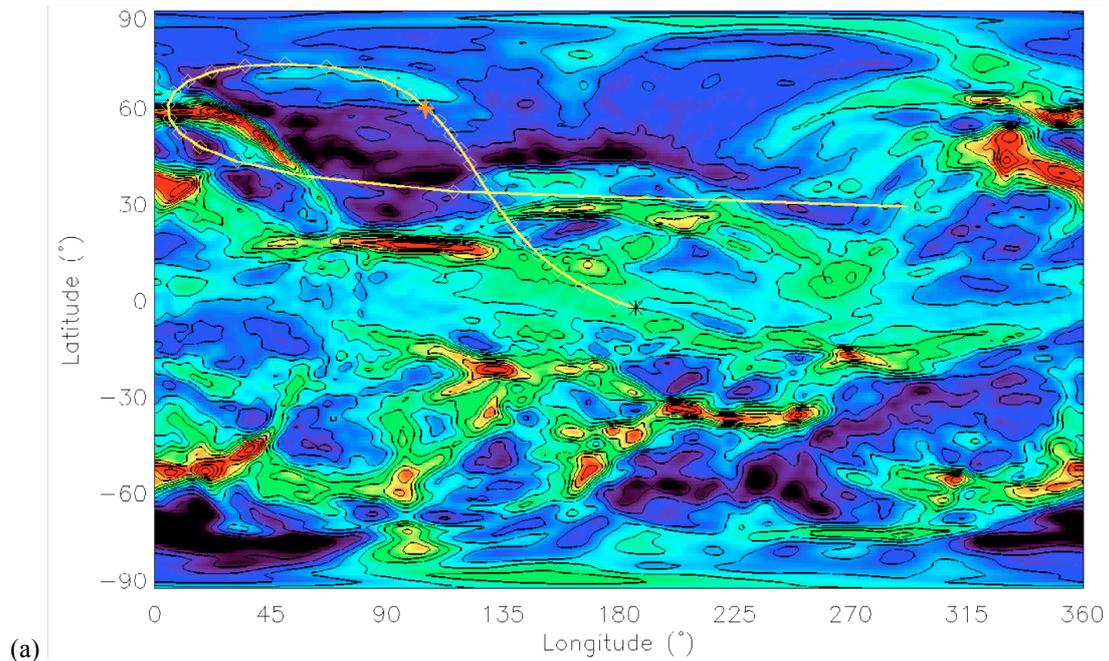

(a)

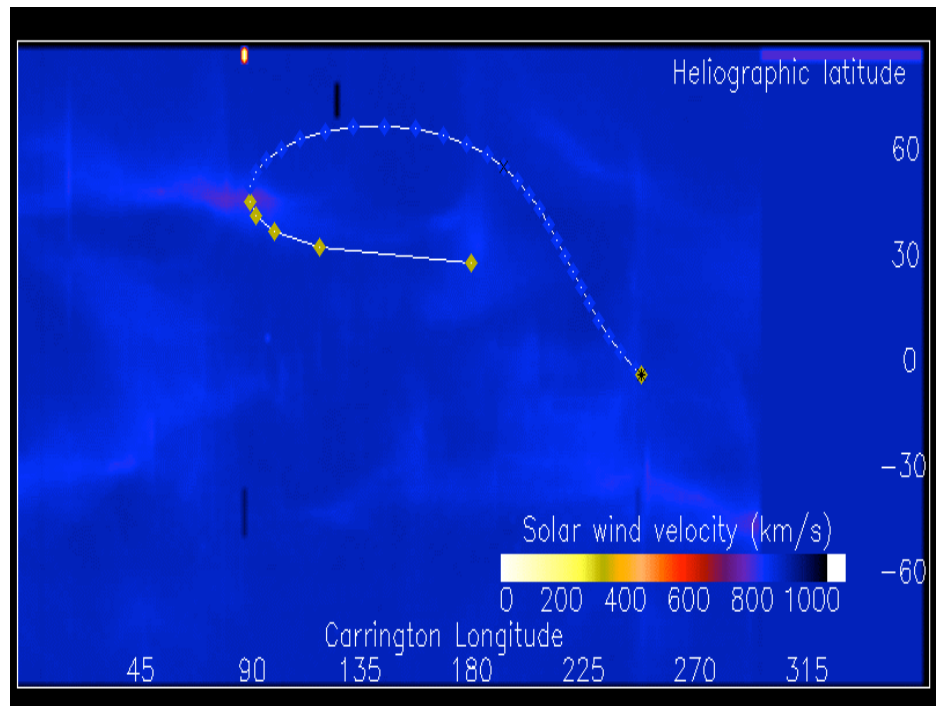

(b)

**Figure 1.** Ballistic raypath maps for source J0319+415 on 15 May 2002 (a) and 12 May 2004 (b). Both raypaths are ballistically back-mapped onto the corona, down to an altitude of 2.5 R$_\odot$, The map in 1(a), and Figure 2 were created using Fourier Back-projection Tomography of LASCO Coronagraph data (Morgan, Habbal, and Lugaz, 2009). In 1(a) denser coronal structures, such as streamers, are coloured red, whilst more rarefied structures are blue or black. The distance between each open diamond on the raypath is equivalent to 5° of separation and the orange star denotes the P-Point. In 1(b) lighter regions of the corona correspond to denser structures whilst darker regions correspond to rarefied regions. The diamonds along the raypath again denote 5°, whilst the colour corresponds to the solar wind velocity scale and indicates where fast or slow is crossing the raypath. The P-Point is indicated by the black X and Earth is indicated by the black asterisk.

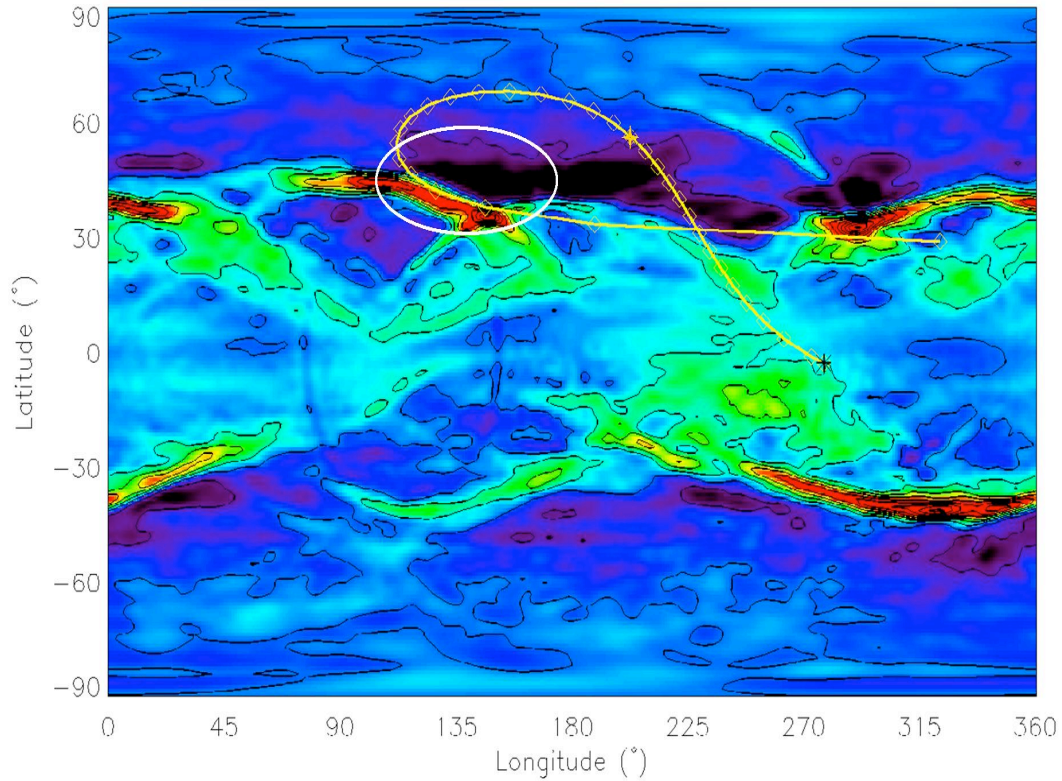

**Figure 2.** The raypath projection, again down to 2.5 $R_S$, for the 13 May 2005 observations shows a significant proportion of the raypath overlying a fast/slow stream boundary (circled), implying the presence of a stream interaction region (SIR), as described by Bisi *et al.*, 2010. The open diamonds again denote 5° of separation along the raypath and the Earth is indicated by the black asterisk. The density of coronal structures scales from red (dense) to black (rarefied). The P-Point, where most of the observed IPS scattering events occur, is indicated by the orange star and it lies some 65° Earthward of the SIR boundary in unperturbed northern fast solar wind.

The raypaths for the observations of J0521+166 on 10 and 11 June 2006 (Figure 3 parts a and b respectively) lay overwhelmingly above the southern polar hole and these observations can be interpreted as sampling unperturbed flow from very high southern latitudes.

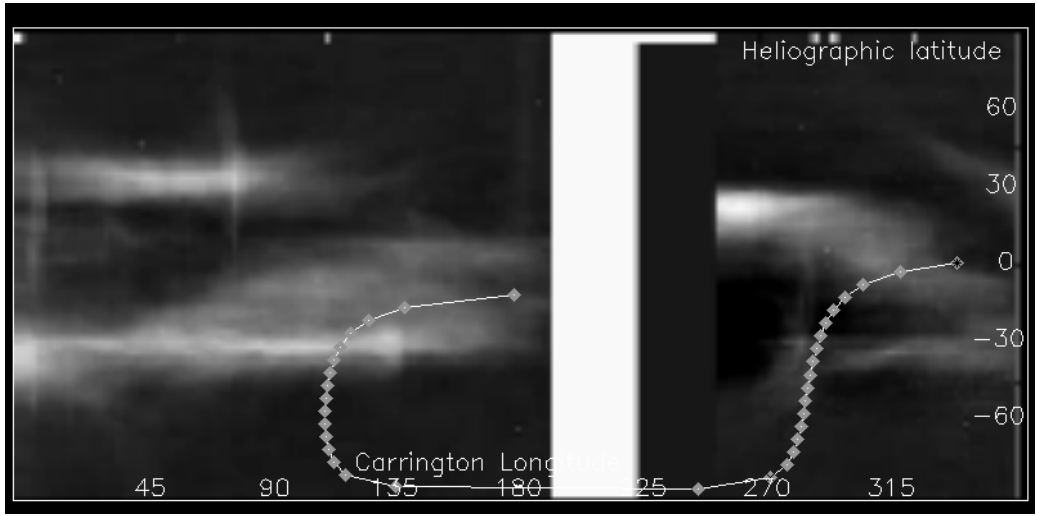

(a)

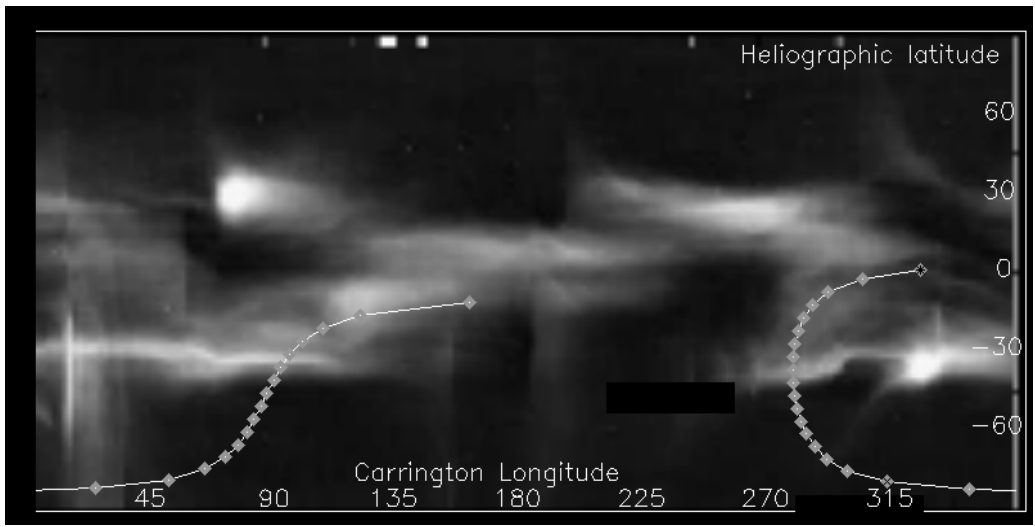

(b)

**Figure 3.** The ballistic raypath projections for 10 and 11 June 2006 (a and b respectively) show the raypaths ballistically back-mapped down to 2.5 R$_\odot$ onto LASCO synoptic coronal maps. Lighter regions indicate denser coronal structures whilst darker regions indicate rarefied regions. Each diamond corresponds to 5° of separation along the raypath. The Earth is indicated by the black asterisk and the P-Point by the black X. The large vertical white feature in 3(a) is a data gap. As can be seen, the raypaths on both days are largely immersed in high southern latitude fast solar wind.

## 3. Analysis

The solar wind flow direction was determined based on the observing geometry for which the highest degree of correlation between the scintillation patterns observed at two different observing sites was recorded. This geometrical approach is described first by Moran *et al.* (1998) and is robust, however employment of this method requires careful correction for the motion of the Earth in the time between a given solar wind feature crossing two different simultaneous raypaths to the same radio source from different receivers. The Earth's orbital motion has the effect of foreshortening both $B_{par}$ and $B_{perp}$ (the parallel and perpendicular baselines respectively) (see Figure 4) and must be corrected before considering the variation of maximum cross-correlation with baseline geometry.

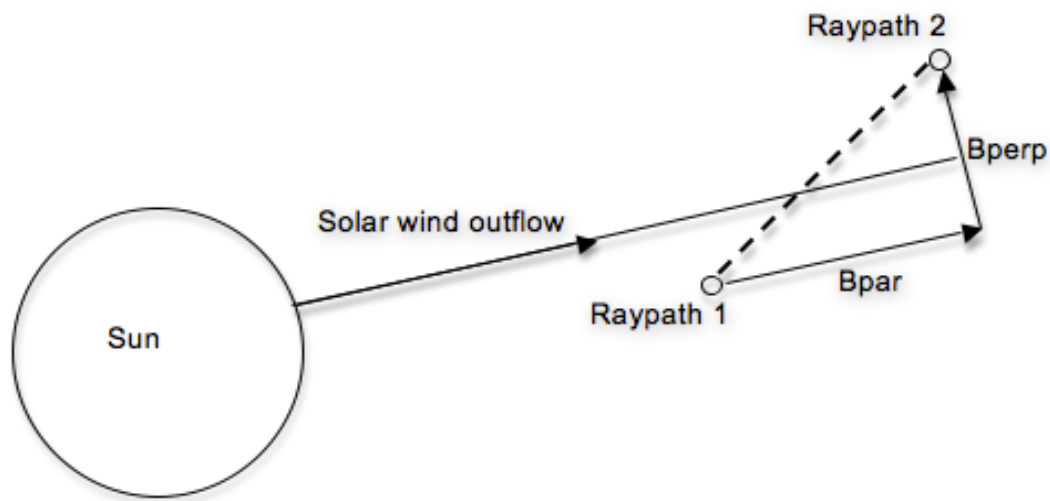

**Figure 4.** A schematic view from the radio source on to the Sun-Earth line showing the origin position of the two raypaths (not to scale) for each antenna. As the Earth rotates during an observation the baseline between raypath 1 and raypath 2 will rotate, thereby changing the lengths of the perpendicular ($B_{perp}$) and parallel ($B_{par}$) components of the baseline. In a truly radial solar wind, the maximum cross correlation between the observed IPS from each raypath occurs when $B_{perp} = 0$.

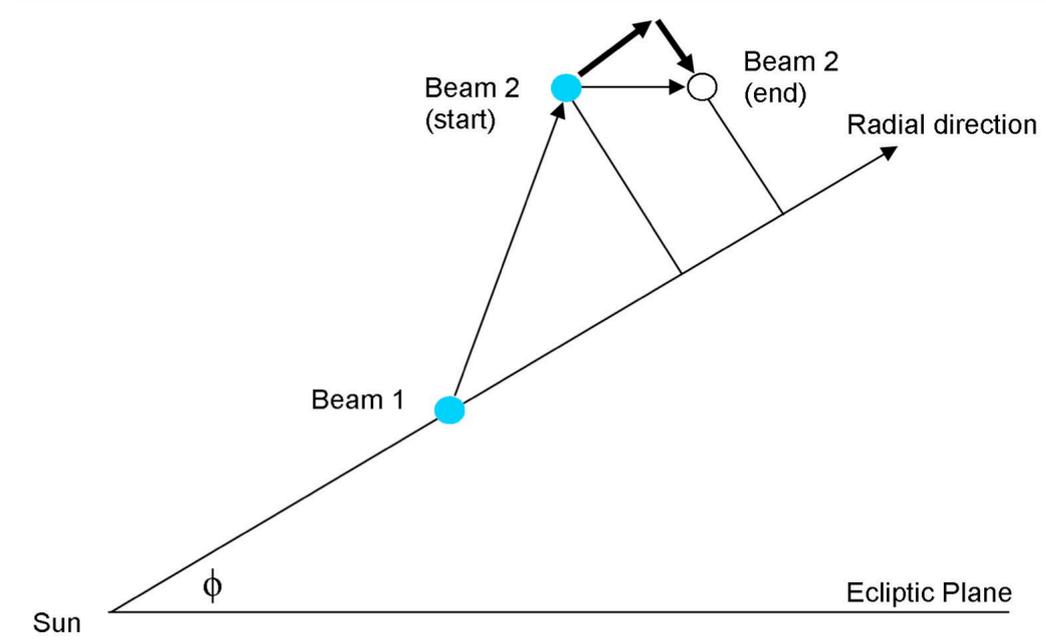

**Figure 5.** During the passage of a solar wind feature between the two raypaths, the Earth will have moved by some non-negligible distance (Beam 2 start to end), which will affect the time at which the peak of cross-correlation is observed and hence will impact the apparent flow direction. The arrow pointing from Beam 1 to Beam 2 (start) indicates the flow direction of the solar wind, which in this case is non-radial (Image adapted from Moran *et al.*, 1998).

In our analysis, the data from an observation are binned into 10-minute sequential time bins. The results from the equations given here are then added to the mean values for $B_{par}$ and $B_{perp}$ in each data bin to account for the motion of Earth using the following equations:

$$B_{par}(corrected) = B_{par}(initial) + V_\varepsilon \tau \cos\phi \sin\theta$$

$$B_{perp}(corrected) = B_{perp}(initial) + V_\varepsilon \tau \sin\phi \sin\theta$$

Where:

$V_\varepsilon$ = orbital velocity of Earth at the time of the observation (typically 29 km s$^{-1}$);

$\tau$ = time difference in seconds, between corresponding scintillation patterns being observed at each receiver;

$\theta$ = angle Earth-Sun-IPS P-point; and

$\phi$ = heliocentric latitude of the radio source as viewed from Earth at the time of the observation. The angles $\theta$ and $\phi$, are illustrated in Figure 6.

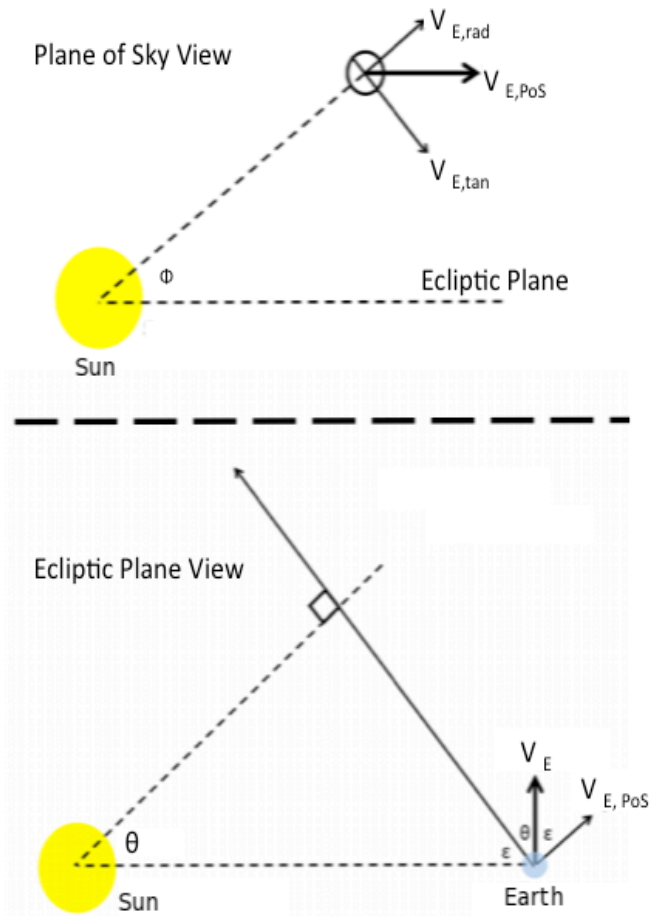

**Figure 6.** The angles θ and φ, in the ecliptic plane view (bottom) and PoS view (top). The motion of Earth must be corrected for during long-duration ELB observations of IPS. The parameter $V_{E,rad}$ is the radial velocity of the Earth. $V_{E,PoS}$ is the Earth's orbital velocity projected onto the PoS. $V_{E,tan}$ is the tangential component of the Earth's orbital velocity.

Figure 7 shows the signing convention we use to establish to which meridional direction (*i.e.* polewards or equatorwards) any deviation in fast solar wind flow is directed. This is defined as the cross-product between the lines of sight to the Sun and to the radio source, in that order. So, for example, source J0319+415 was situated off the North-Eastern quadrant of the Sun in all observing campaigns and hence any equatorwards deviation of solar wind flow would be indicated by a negative off-radial angle, whereas polewards deviation would be indicated by a positive off-radial angle. All four quadrants are clearly indicated in the figure.

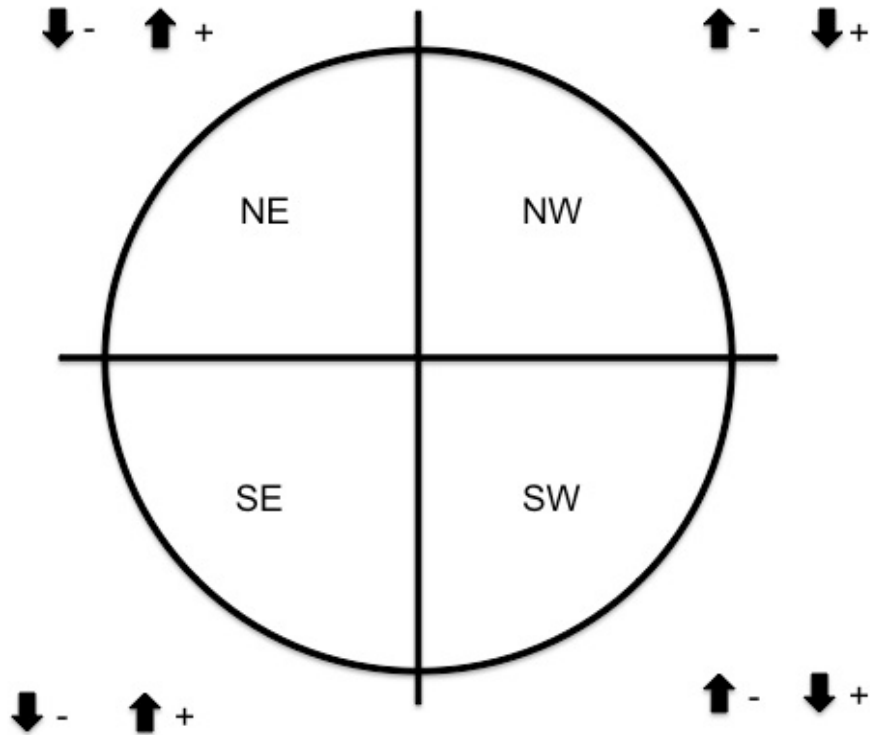

**Figure 7**. The sign convention used to define off-radial (meridional) flow direction in this study is shown here based on heliographic coordinates. Source J0319+415 lay in the north-eastern (NE) quadrant in all observation campaigns, whereas J0521+166 lay in the south-eastern quadrant (SE).

## 4. Results

Results from observations of the appearance of maximum correlation between different receiver-to-receiver baselines are shown in Figures 8-13. Each point in these plots represents the receiver-to-receiver correlation observed for a given 10-minute data collection-and-integration period, and its corresponding off-radial (meridional) flow direction using the geometrical method described earlier in Section 3. The plots include data from campaigns in 2002, 2004, 2005 (Bisi, 2006; Bisi *et al*., 2005; 2007), and 2006 (Dorrian, 2009).

Some filtering of data has been performed to reduce ambiguity in the observations. Because of the constantly-changing geometry of the projected baselines due to Earth's rotation and orbital motion, cross-correlated IPS signals will produce meaningful results on different baseline combinations at differing times. Furthermore, with the use of two-site ELB IPS, the two raypath positions at each end of the baseline will effectively "orbit" each other quite rapidly compared to observations using shorter baselines. This somewhat reduces the time resolution with which we can observe changing correlation height over the course of the observing run and it is therefore possible that the peak of correlation occurs very close to the start or end of an observation, or even beyond them (for certain baseline combinations). In the type of plots shown in Figures 8-13, this would lead to ambiguity in identifying the correlation peak as the correlation height beyond the start and end times of the observing run was not known. Any observations which displayed this characteristic have been rejected.

In addition, only observations with at least three data points in the series have been included, again to enable unambiguous identification of a correlation peak. Any data series with fewer than three points have been rejected. In all cases, as can be seen in the plots, the highest levels of cross-correlation are typically concentrated around 3-4° equatorwards of the radial flow axis (see discussion in Section 5), based on the signing convention defined in Figure 7 and discussed in the previous section.

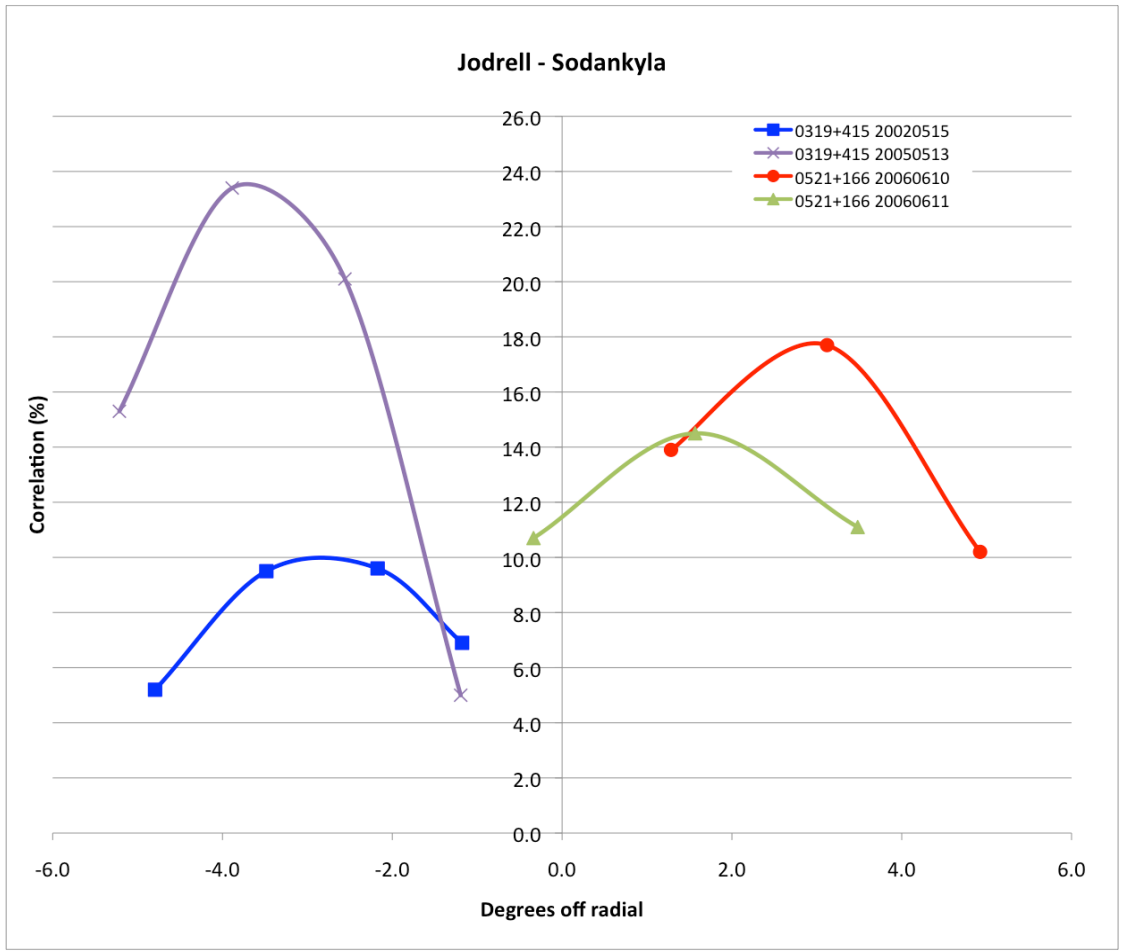

**Figure 8.** Jodrell-Sodankylä baselines showing recorded IPS correlation for radio sources J0319+415 and J0521+166, respectively, from 2002 to 2006. Source J0319+415 lay off the north-eastern (NE) quadrant of the Sun (as viewed from Earth) at high latitudes in 2002 and 2005. J0521+166 lay of the south-eastern (SE) quadrant, again at a high latitude. As data sets from two days are available from 2006, each have been stated explicitly. Given the signing convention defined in Figure 7, a -ve value for off-radial (meridional) flow in the NE hemisphere indicates an equatorwards deflection of the solar wind stream, whereas in the SE quadrant, such a deflection would be indicated by a +ve angle. Equatorwards deflection of the fast solar wind, of the order of 3° is therefore observed here in both hemispheres, and at different times in the solar cycle.

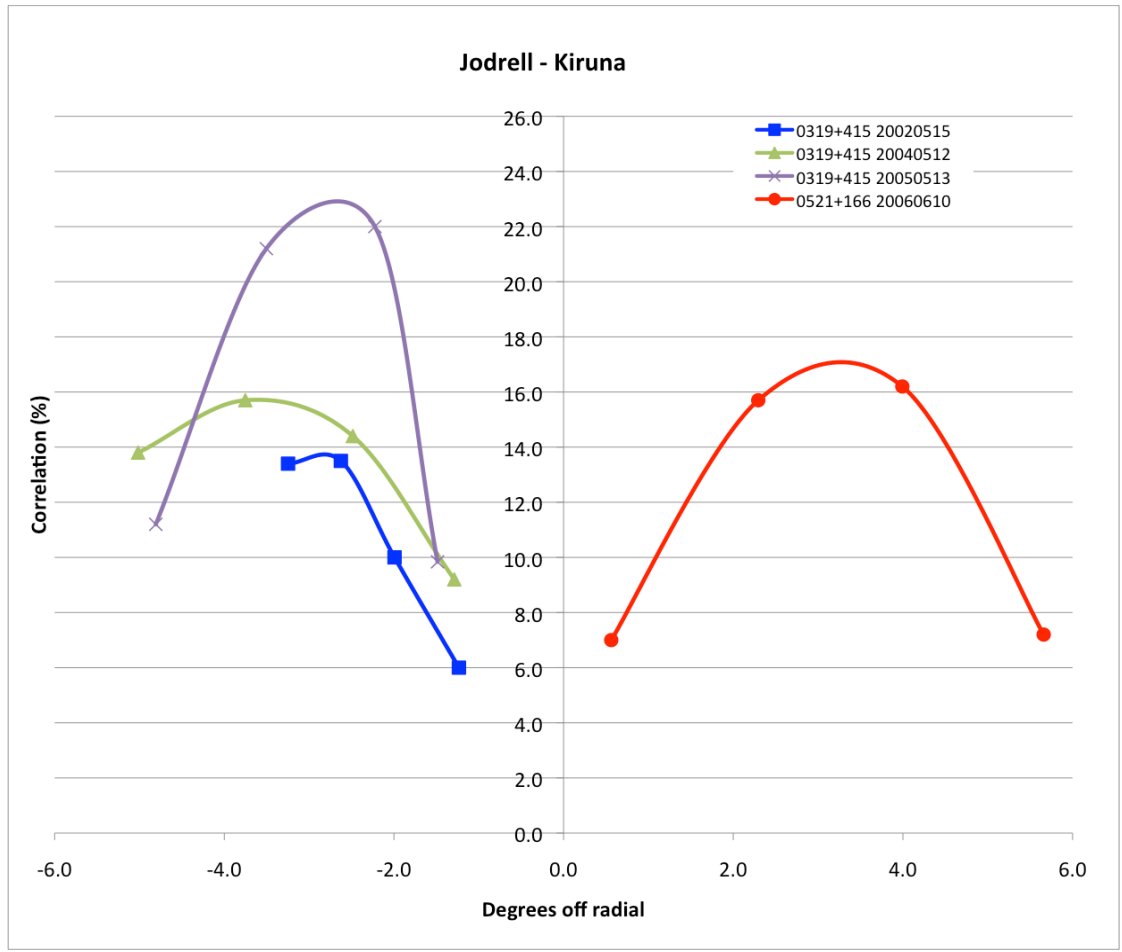

**Figure 9.** Jodrell-Kiruna baselines showing recorded IPS correlation observed on source J0319+415 and J0521+166, as in Figure 8 for 2002 to 2006. In all cases an equatorwards deviation of ~3° is observed in general.

Figures 10-13 show data from the Cambridge-Tromsø, Cambridge-Kiruna, Cambridge-Sodankylä, and Knockin-Kiruna baseline pairs respectively. Again, in all cases, a negative angle deviation is observed for fast solar wind streams in the North-Eastern quadrant (on source J0319+415) whilst positive angle deviations are observed for all sampled solar wind flow in the South-Eastern quadrant (on source J0521+166). Based on the signing convention defined in Figure 7, this indicates that fast solar wind flow is deviating equatorwards in all cases.

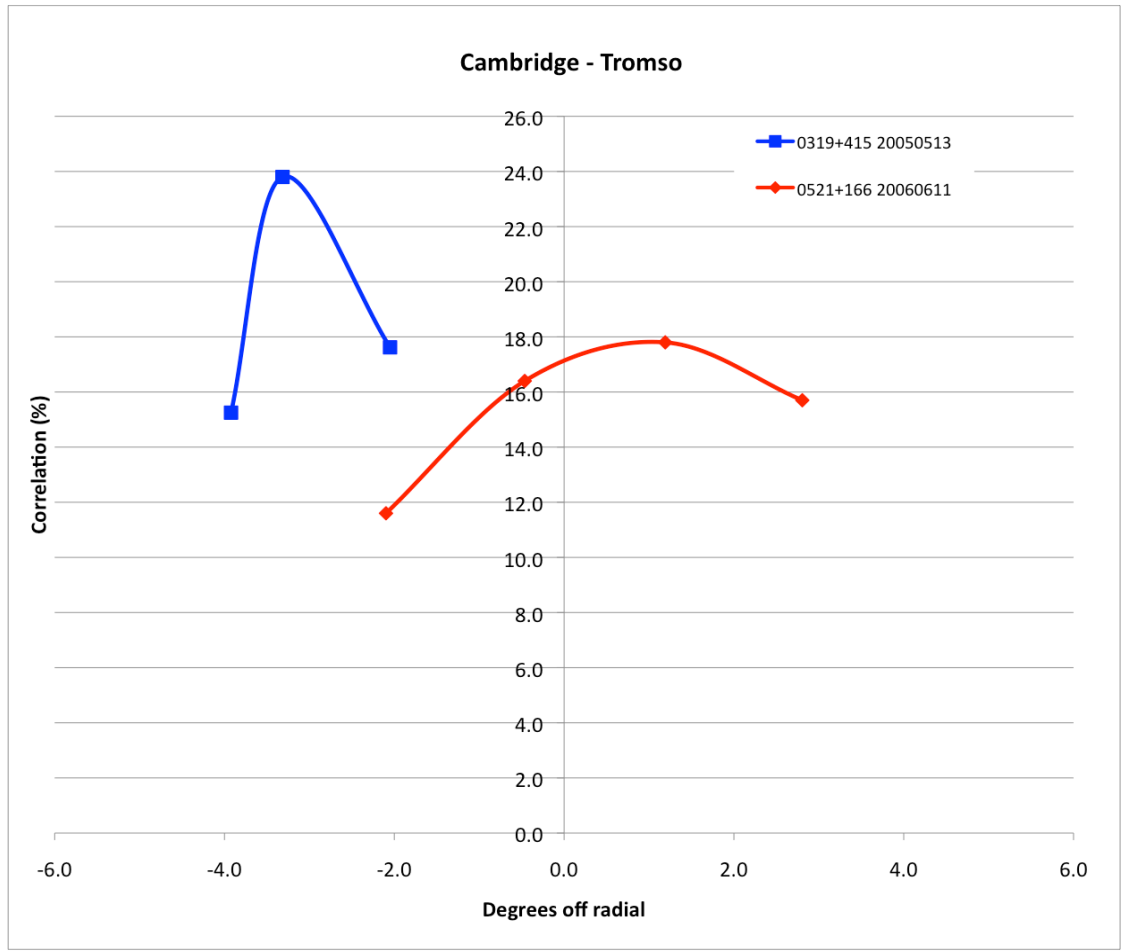

**Figure 10**. Cambridge-Tromsø baseline data from 13 May 2005 and 11 June 2006 using radio sources J0319+415 and J0521+166 respectively. Equatorwards deviation in fast solar wind flow is again of the order of 3° (as in Figures 8 and 9) whilst observing scintillation on source J0319+415. Similar deviation is present, but less pronounced, in the 11 June 2006 observation using source J0521+166.

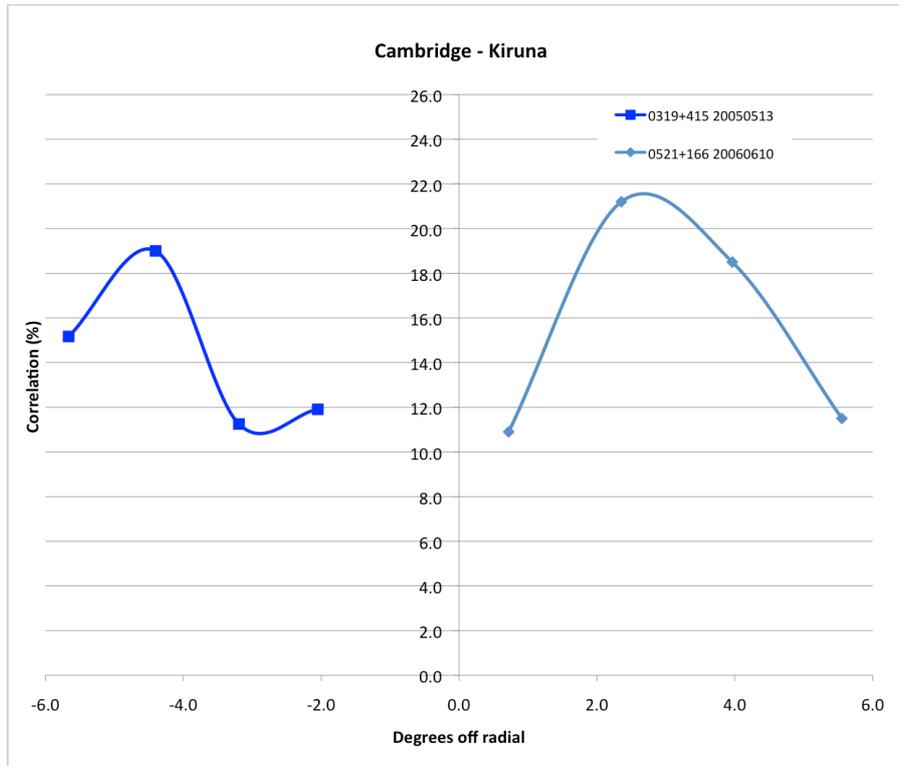

**Figure 11.** Cambridge-Kiruna baseline data from 2005 and 2006. Equatorwards deviation of ~4° in fast solar wind flow in the NE quadrant is observed with scintillations of source J0319+415 in 2005 and in the SE quadrant on source J0521+166. Again, this is consistent with the other previously-shown baseline pairs.

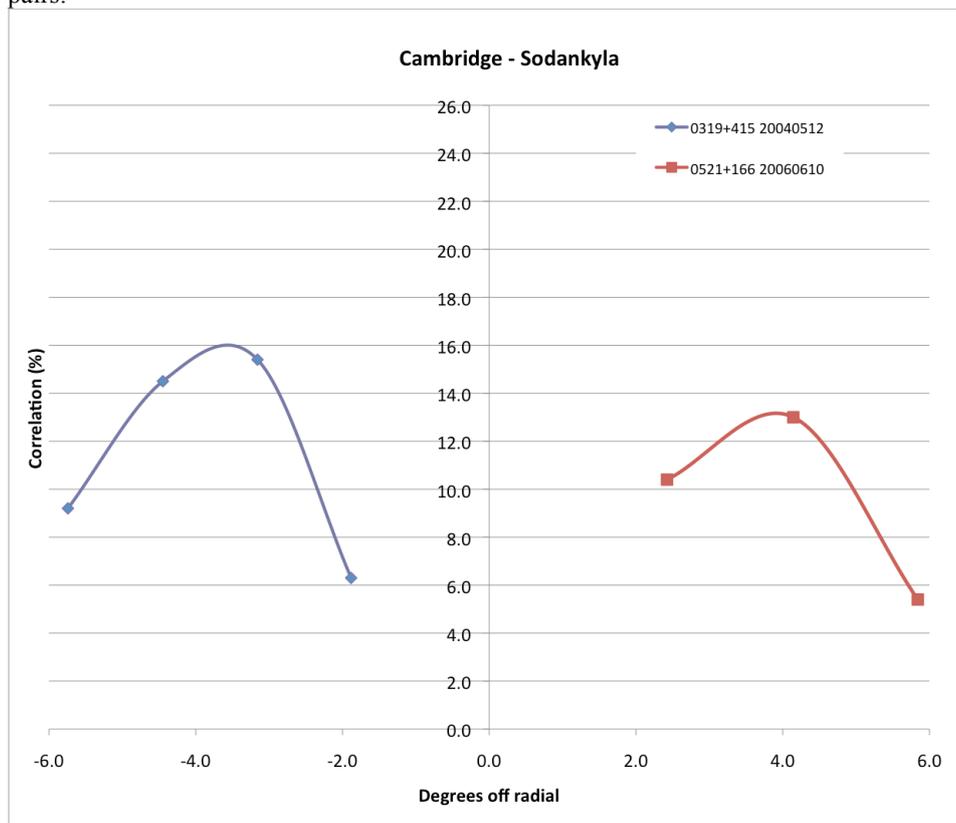

**Figure 12.** Cambridge-Sodankylä baseline results from 2004 and 2006, again using radio sources J0319+415 and J0521+166. Equatorwards deviation of ~4° in fast solar wind flow is seen in both cases.

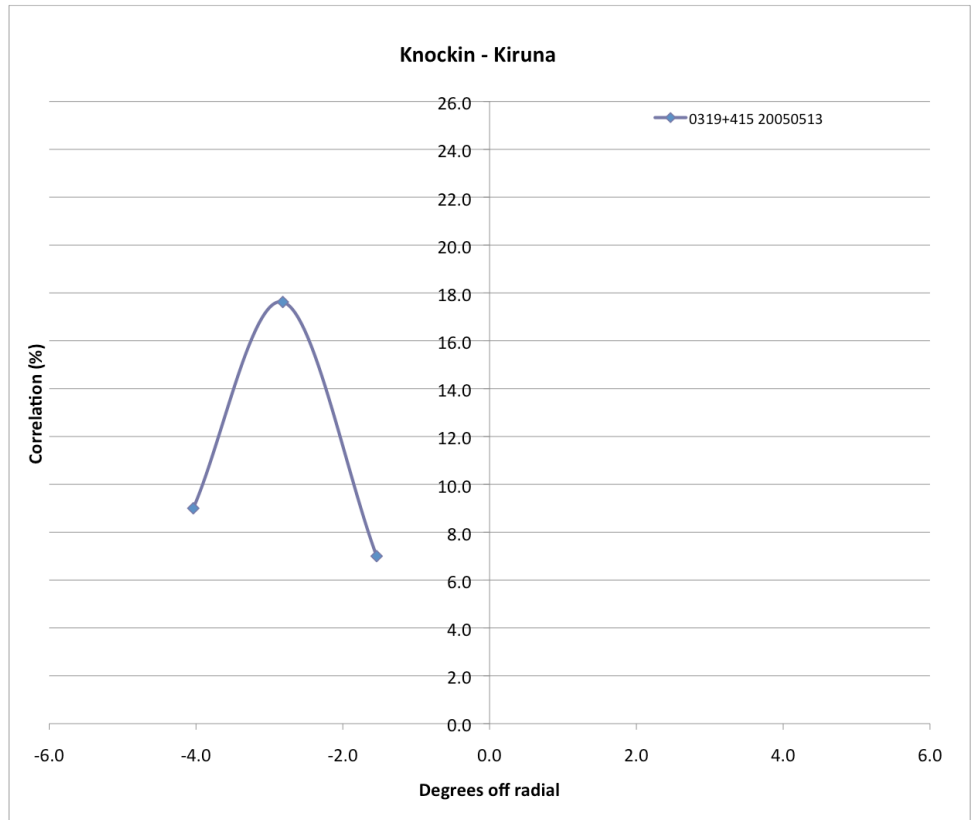

**Figure 13**. Knockin-Kiruna baseline with source J0319+415 on 13 May 2005. As in all previous cases using this radio source, equatorwards deviation of fast solar wind of ~3° is observed.

## 5. Discussion and Summary

The technique of observing ELB IPS has been used to demonstrate flow characteristics of the fast solar wind over both hemispheres of the Sun, during unperturbed flow. Joint observations made with two different radio sources at different PoS distances from the Sun, with multiple baseline antenna pairs and at different times in the solar cycle, have demonstrated that the fast solar wind exhibits measureable equatorwards flow characteristics, in so far as this technique can determine.

The Jodrell Mk. 1 'Lovell' Telescope provided a strong signal-to-noise (S/N) performance throughout the observations enabling more reliable data returns from observing runs involving this antenna as one of the baseline pairs. This is evident from the larger number of baseline combinations shown involving the Lovell (Figures 8 & 9). Higher S/N ratios enable the shortening of the time resolution between data bins. More accurate determinations of the time of peak correlation are highly desirable in a study of this kind and hence the future availability of antennas with strong S/N performances would prove highly valuable. Further observation campaigns using additional radio sources, especially in the NW and SW quadrants are also desirable, as flow characteristics in these quadrants have not yet been sampled, although we expect them to exhibit the same behaviour.

Typical peaks in cross correlation throughout these series of observing campaigns correspond to off-radial (meridional) flow angles of some 3-4° in the equatorwards direction. Accounting for the limitations in sensitivity of the technique used here, these observations place a lower constraint on the length of radial baseline ($B_{par}$) required in order to detect any non-radial deviation in fast solar wind flow.

Thus, in summary, we therefore suggest that the hypothesis advocated in the Parker solar wind model (*e.g.* Parker, 1963; Neugebauer and von Steiger, 2003), that unperturbed fast solar wind follows a primarily radial flow direction, is not strictly valid. We emphasise the use of multiple and, in some cases, completely independent baseline pairs (*e.g.* Jodrell-Sodankylä and Cambridge-Kiruna), during different times in the declining phase of Solar Cycle 23, in making these consistent observations of equatorwards deviations in high latitude fast solar wind flow.

**Acknowledgements**  We would like to thank the directors and staff of both the *European Incoherent SCATter* (EISCAT) radar consortium and the *Multi-Element Radio Linked Interferometer Network* (MERLIN). EISCAT is supported by the scientific research councils of Finland, France, Germany, Japan, Norway, Sweden, the UK, and China. MERLIN is supported by the UK Science and Technology Facilities Council (STFC). Four of us (Dorrian, Breen, Fallows, and Bisi) were also supported by STFC throughout the majority of this research.

# References

Armstrong, J.W., Coles., W.A.: 1972, Analysis of three station interplanetary scintillation, *J. Geophys. Res.*, **77**, 4602 - 4610.

Bisi, M.M.: 2006, Interplanetary Scintillation Studies of the Large-Scale Structure of the Solar Wind. Ph.D. Thesis, The University of Wales, Aberystwyth (http://cadair.aber.ac.uk/dspace/handle/2160/4064).

Bisi, M.M., Breen, A.R., Fallows, R.A., Thomasson, P., Jones, R.A., Wannberg, G.: 2005, Combined EISCAT/ESR/MERLIN Interplanetary Scintillation Observations of the Solar Wind, ESA, *Proceedings of the Solar Wind 11 / SOHO 16*, 593-596.

Bisi, M.M., Fallows, R.A., Breen, A.B., Habbal, S.R., Jones, R.A.: 2007, Large scale structure of the fast solar wind. *J. Geophys. Res.* **112**(A6), A06101.

Bisi, M.M., Fallows, R.A., Breen, A.R., O'Neill, I.J.: 2010, Interplanetary Scintillation Observations of Stream Interaction Regions in the Solar Wind. *Solar Phys.* **261**, 149-172.

Bourgois, G., Daigne, G., Coles, W.A., Silen, J., Turunen, T., Williams, P.J.S.: 1985, Measurements of the solar wind velocity with EISCAT, *Astron. Astrophys.*, **144**, 452-462.

Breen, A.R.; Coles W.A., Grall, R., Lovhaug, U.–P., Markkanen, J., Misawa, H., Williams, P. J. S.: 1996, EISCAT measurements of interplanetary scintillation, *J. Atm. Terr. Phys.*, **58**, 507-519.

Breen. A.R., Riley, P., Lazarus, R.J., Canals, A., Fallows., R.A., Linker, J., Mikic, Z.: 2002, The solar wind at solar maximum: comparisons of EISCAT IPS and in situ observations, *Ann. Geophys.* **20**, 1291-1309.

Breen. A.R., Fallows, R.A., Bisi, M.M., Thomasson, P., Jordan, C.A., Wannberg, G., Jones, R.A.: 2006, Extremely long baseline interplanetary scintillation measurements of solar wind velocity, *J. Geophys. Res.* **111**, A08104.

Breen A.R., Fallows, R.A., Bisi, M.M., Jones, R.A., Jackson, B.V., Kojima, M., Dorrian, G.D., Middleton, H.R., Thomasson, P., and Wannberg, G.: 2008, The solar eruption of 2005 May 13 and its effects: Long-baseline interplanetary scintillation observations of the Earth-directed coronal mass ejection, *Astrophys. J. Lett.*, **683**, L79-L82.

Coles W.A.: 1995, Interplanetary Scintillation Observations of the High-Latitude Solar Wind, *Space Sci. Rev.*, **72**, 211-222

Dennison, P.A., Hewish, A.: 1967, The Solar Wind outside the Plane of the Ecliptic, *Nature*, **213**, 5074, 343-346.

Dorrian, G.D.: 2009, Large Scale 3-Dimensional Structure of the Solar Wind. Ph.D. Thesis, Aberystwyth University (http://cadair.aber.ac.uk/dspace/handle/2160/3125).

Fallows, R.A., Williams, P.J.S., Breen, A.R.: 2002, EISCAT measurements of solar wind velocity and the associated level of interplanetary scintillation, *Ann. Geophys.*, **20**, 1279-1289.

Fallows, R.A., Breen, A.R., Bisi, M.M., Jones, R.A., Wannberg, G.: 2006, Dual-frequency interplanetary scintillation observations of the solar wind, *Geophys. Res. Lett.* **33**, L11106.

Fallows, R.A., Breen, A.R., Dorrian, G.D.: 2008, Developments in the use of EISCAT for interplanetary scintillation, *Ann. Geophys.*, **26**, 2229-2236.

Grall, R.R., Coles, W.A., Klinglesmith, M.T.: 1996, Observations of the solar wind near the Sun, *AIP CS-382*, 108-108.

Hewish, A., Scott., P.F., Wills, D.: 1964, Interplanetary Scintillation of small diameter radio sources, *Nature*, **203**, 1214-1217.

Moran, P.J., Breen, A.B., Varley, C.A., Williams P.J.S., Wilkinson W.P., Markkanen, J.: 1998, Measurements of the direction of the solar wind using interplanetary scintillation, *Ann. Geophys.*, **16**, 1259-1264.

Morgan, H., Habbal, S.R., Lugaz, N.: 2009, Mapping the Structure of the Corona Using Fourier Backprojection Tomography, *Astrophys. J.*, **690**, 1119-1129.

Neugebauer, M., von Steiger, R.: 2003, The Solar Wind, in J.A.M. Bleeker, J. Geiss, M.C.E. Huber, eds., *The Century of Space Science,* chapter 47, 1115–1140, Kluwer Academic Publishers.

Parker, E.N.: 1963, Interplanetary Dynamic Processes. Interscience Publishers, New York.

Rishbeth, H., Williams, P.J.S.: 1985, The EISCAT ionospheric radar - The system and its early results, Royal Astronomical Society, Quarterly Journal, **26**, 478-512.

Thomasson, P.:1986, MERLIN, Royal Astronomical Society, Quarterly Journal, **27**, 413-431.